\documentclass[twocolumn,amsmath,amssymb]{revtex4}
 \begin{document}
                                                                                
\title{Gobal entanglement and double occupancy in many-electron states}
\author{ V. Subrahmanyam}
\address{ Department of Physics, Indian Institute of Technology, Kanpur-208016, India}
\date{\today}
\begin{abstract}
The entanglement in many-electron states is investigated using a global
entanglement measure, viz. average site mixedness. We have examined metallic
states  of noninteracting electrons, Nagaoka and Gutzwiller states of strongly-correlated electrons,
and superconducting states.
Uncorrelated metallic states at half filling seem to maximize entanglement, as these states
optimize the number of holes, the number of
doubly-occupied sites.
Entanglement is calculated explicitly for
Gutzwiller-projected many-electron states in one dimension, which have less entanglement as
double occupancy is inhibited in these states. 
Entanglement in superconducting states, which tend to promote double occupancy,  is calculated
as a function of the energy gap, and found to be  lower than the metallic state entanglement.
There is a possibility of  a regime with a nonzero single-site concurrence depending on the energy gap.
\end{abstract}
\maketitle

Quantum entanglement is perceived as a resource for quantum communication
and information processing, and has emerged over the last few years
as a major research area in
various diverse fields such as physics, mathematics, chemistry, electrical
engineering and computer science\cite{Nielsen, Dagmar,Benenti}. There has been a wide-spread investigation
of entanglement properties of the ground states and excited states of 
well-studied quantum spin models of the condensed matter physics like the
Heisenberg-XY models, the transverse-field Ising model\cite{Zanardi, Wang, Arul, Subra, Osborne,Arul2004}. Also, there has been an intense study of the dynamics of entanglement of 
these models, both from numerical and analytical approaches.

The spin-only states have several advantages; many entanglement measures  have been successfully used to characterize and quantify the entanglement of
many-spin states of lattices. Pair-wise entanglement or concurrence\cite{Wootters}, which
measures the entanglement of a given pair of spins in a many-spin state,
has been widely studied both numerically and analytically\cite{Zanardi, Arul,Subra}. The global entanglement measure\cite{MeyerWallach}, which quantifies how entanglement is distributed and shared over
various spatial parts of the lattice, has been studied.  The reduced fidelity susceptibility measure
has been studied to investigate the critical properties of spin systems\cite{Fidel}.
In contrast, the entanglement properties of general many-electron states, which combine
both spin and orbital degrees of freedom, have not been studied extensively.
In this article we will use a 
generalization of the global entanglement measure, 
viz. the average site mixedness, to quantify the entanglement between
spatial parts of many-electron states on lattices.
  
Let us consider electrons either localized or hopping around on $N$ lattice
sites. The details of the interactions determine the exact many-electron ground state that we may like to
study. Since the entanglement properties can be discussed entirely through
the many-body state itself, we will refer to the interactions only through
the structure of the state in this article. 
There are four states per site, implying a basis of $4^N$ states
for the $N-$site lattice. A convenient basis states for any site is given
by 
$|0\rangle,|\uparrow\rangle,|\downarrow\rangle,|\uparrow\downarrow\rangle$; which
correspond to no occupancy (or a hole),  up-spin electron occupancy, down-spin
electron occupancy, and double occupancy respectively. 
A general  pure  many-electron state for $N$ sites can be written as,
\begin{equation}
|\psi\rangle=\sum_{\lbrace a_i,..a_N \rbrace} \phi_{a_1,a_2,...a_N} |a_1,a_2,...a_N \rangle,
\end{equation}
where $a_i$ labels the basis states of site $i$ listed above.
This state is characterized by $4^N$ complex-number amplitudes $\lbrace\phi_{a_1,a_2..}\rbrace$ for
the basis states. 
The entanglement distribution in the  state can be
investigated using various reduced density matrices (RDM) constructed through
partial traces over the density matrix of the system $\rho=|\psi\rangle \langle \psi|$.
In general a RDM of a subsystem would correspond to a mixed state, implying
an entropy for the subsystem, and a bipartite entanglement between the 
subsystem and the rest of the system, and also some amount of multi-party 
entanglement between the various sites of the subsystem.  It is quite an uphill task to characterize and quantify the entanglement  distribution in various partitions of the system.

In the context of many-electron states even two-site RMDs are difficult to
work with and unravel the pair-wise entanglement, as the RDM is now
16-dimensional since no occupancy and double occupancy can occur.  Here we will use  the
single-site RDMs, and use a generalization of the
global entanglement measure\cite{MeyerWallach}, that has been used extensively in spin systems.
The entanglement measure is given in terms of  the single-site RDM  $\rho_l$ as,
\begin{equation}
\varepsilon(\psi)\equiv {4\over 3 N} \sum_{l=1}^N ( 1 - Tr \rho_l^2).
\end{equation} 
In the above, the pre-factor is so chosen that if each site density matrix
is maximally mixed (corresponding to all four eigenvalues being equal to 1/4),
the entanglement measure equals unity. Since, 1- $Tr \rho_l^2$ would be zero
for a pure single-site RDM, the above measures average site mixedness or how 
entangled is a site with the rest of the system on an average. 
Though we are working with
strictly local, single-site RDMs, the averaging process makes it a global 
measure.  If a large number of sites with nonzero measure are entangled with 
other sites, we get a nonzero entanglement. 

Let us consider a quantum state  with $N_e=nN$ number of electrons, where $n$ is the
electron density. The state is further characterized by the densities of up and
down spin electrons, with $N_\uparrow=n_\uparrow N, N_\downarrow=n_\downarrow N$
as the up and down spin electron numbers respectively. 
That is, we consider
a quantum state with conserved densities of up and down-spin electrons. 
Because of the conserved electron densities, the
single-site RDM of site $i$ has a diagonal structure, using the
site basis states as unoccupied,doubly-occupied, up-spin-only occupied, down-spin-only occupied states, it is given as 
\begin{equation}
\rho_i=\left(
\begin{array}{cccc}
1-n+d  & 0&0&0\\
0 & d&0&0\\
0 & 0&n_{\uparrow}-d&0\\
0 &0& 0&n_{\downarrow}-d
\end{array}\right).
\end{equation}
In the above $d$ is the probability of double occupancy, which related to the total number of
doubly-occupied sites in the state as $Nd\equiv D=<\sum n_{i\uparrow}n_{i\downarrow}\rangle$.

From the structure of the single-site RDM, we can infer a hierarchy of
entanglement in many-electron states. 
 The spin-only states $\{|\psi_1\rangle\}$,  with evert site being occupied by
an electron with either up or down spin, correspond to the half-filled case
of strongly-correlated electron states.  Here, we have $d=0, n=1$,
i.e. no double occupancy and no holes, effectively reducing
the local Hilbert space to two, and the single-site
RDM has only two eigenvalues. The maximum entanglement occurs when both
these eigenvalues are same, i.e. $n_\uparrow=n_\downarrow=1/2$. Thus we have from Eq.3, 
\begin{equation}
\varepsilon_{max}(\psi_1)={2\over 3}.
\end{equation}
It should be pointed out that there will be many spin states with this
entanglement $\varepsilon=2/3$. In fact, all many-electron states with the
total spin zero, no double occupancy, and no holes will satisfy this.

The second category comprises of states $\{|\psi_2\rangle\}$ 
with no doubly-occupied sites but a nonzero fraction of holes, $d=0$,$n<1$. 
Since the site Hilbert space now is three dimensional, we have three nonzero
eigenvalues of the single-site RDM. These states correspond to low-lying states of a large-U
Hubbard model, or very strongly-correlated electron states with less than half filling electron density. Here, the maximum entanglement occurs
for $1-n=n_\uparrow=n_\downarrow=1/3$, with the eigenvalues of RDM being equal to
1/3, we have
\begin{equation}
\varepsilon_{max}(\psi_2)={8\over 9}.
\end{equation}
Thus existence of holes increases the entanglement.
%

Finally,  let us consider states with doubly-occupied sites
as well. Maximum entanaglement occurs if all the four eigenvalues of the
RDM shown above are all equal, i.e. $d=1/4,n_\uparrow=n_\downarrow=1/2$. This corresponds to the half-filling case, except that doubly-occupied
sites are allowed. Since the local Hilbert space dimension has been maximized
to four, this gives the maximum entanglement, a global maximum, 
\begin{equation}
\varepsilon_{max}(\psi)=\varepsilon(d={1\over 4}, n_\uparrow=n_\downarrow={1\over 2})=1.
\end{equation}
In general there will be many such states with $\varepsilon=1$, since the only criteria we used
are the conserved electron densities and the spatial uniformity. However, it will be useful to 
analyze various physical states such as metallic states, strongly-correlated states, superconducting
states etc. Below we will examined these qualitatively different states.

Let us consider the Nagaoka state\cite{Nagaoka},  with $d=0, N_e=N-1$, which is the ground state of $U=\infty$ Hubbard model. This corresponds to a  multiplet with the maximal total spin $S=S_max=N-1/2$. 
However, all the states in the ground state multiplet need not have the
same entanglement. Consider the ground state with $S^z=N-1-l$, i.e. $N_\uparrow=
N-1-l,N_\downarrow=l$. The eigenvalues of the single-site RDM are 
$1/N, 0, 1-(l+1)/N, l/N$, giving an entanglement
$\varepsilon(S^z=N-1-l)={8 (l+1)\over 3 N}(1-{l\over N}).$
It is easy to see that the maximum entanglement occurs for the state with $l=N-1/2$, we have
\begin{equation}
\varepsilon_{max}(\psi_{Nagaoka})={2\over3} (1+{1\over N})^2.
\end{equation}
This illustrates that the presence of a hole increases the entanglement from 2/3, which is the
maximum value for the case of half filling with infinite Hubbard interaction. 

Let us now examine the behavior of the entanglement as a function of $d$, the average density
of the doubly-occupied sites.  A state with an optimal $d$ is the uncorrelated metallic state or
Fermi ground state, which is constructed as a direct product of single-particle momentum states,
$|F\rangle=\prod c^\dag_{k\uparrow} c^\dag_{k\downarrow}|0\rangle$. Each single-particle $k \le k_F$ state is occupied
by two electrons,  and the electron density determines the Fermi wave vector $k_F$. 
In the site basis, the state does not have a direct product structure, thus exhibits entanglement.
 It is easy to see that in the Fermi ground state, the double occupancy is given by $d=n^2/4$, and
 $\varepsilon=1$ for the case of half filling $n=1$.
A popular state that incorporates a double occupancy as the on-site correlation strength is
varied is the Gutzwiller state\cite{Gutzwiller},  given as
 \begin{equation}
|g\rangle= {1\over \sqrt{\cal N}} \prod_{i=1}^N [1-(1-g)n_{i\uparrow}n_{i\downarrow}] |F\rangle,
\end{equation}
where ${\cal N}$ is a normalization factor, $g$ determines the amplitude for a state with double
site occupancy.
It is clear that $g=1$ corresponds to the uncorrelated metallic state,  where as
the other extreme $g=0$ corresponds to no double occupancy or strongest spin
correlations, or to $U=\infty$ Hubbard model ground states.
The double occupancy in the
above state is related to the normalization factor,
\begin{equation}
d(g)={1\over 2N} {\partial \log {\cal N} \over \partial \log{g}},
\end{equation}
which is not easy to calculate in general. In one dimension, it has been calculated by Metzner and Vollhardt\cite{MetznerVollhardt}, we have
\begin{equation}
d(g)={1\over 2} {g^2\over (1-g^2)^2} [-n(1-g^2)-\log (1-n(1-g^2))].
\end{equation}
Using this, the entanglement is easily calculated, and is shown in Fig.1 as a function of the projection
parameter $g$. 
It is seen that
optimal entanglement occurs for the uncorrelated metallic state at $g=1$ at half filling, and  for small electron densities the entanglement at near
$g\sim 0$ can be greater than the corresponding metallic case near $g\sim 1$, though, however,
the actual amount is quite small compared to the global maximum at $n=1, g=1$.
 Thus, the tendency of inhibiting double occupancy, from the optimal metallic value of $d=n^2/4$ ,
decreases in general the site mixedness or the global entanglement. 

We now examine
situations where it is energetically favorable to promote double occupancy,
over and above the optimal metallic value. Superconducting states  fit into this category,
as the underlying attractive interaction mechanism tends to promote superconducting Cooper pairs of
opposite spins and in a zero momentum state, which  is equivalent  to a promoting tendency for site
double occupancy. Starting with the vacuum state $|0\rangle$, the superconducting state can be constructed
as
\begin{figure}[t]
\input{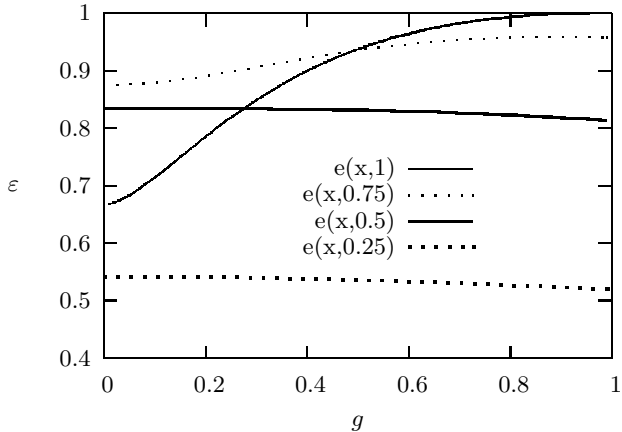}
\caption{ The entanglement in the Gutzwiller state as a function of $g$ the projection amplitude for
  different electron densities. The metallic state is at $g=1$ and the strongly-correlated state at $g=0$.     For small filling fraction the strongly-correlated state has a slightly higher 
entanglement than the uncorrelated state, which itself is quite small compared to the global
maximum of the matallic state at $n=1$.}
 \end{figure}
\begin{equation}
|s\rangle=\prod_{k} [u_k + v_k c^\dag_{k\uparrow} c^\dag_{-k \downarrow}] |0\rangle.
\end{equation}
Here, $c^\dag_{k\uparrow}$ creates an up-spin electron in a  single-particle
plane-wave state with energy $E_k=\hbar^2k^2/2m$.
$v_k$ is the amplitude for a Cooper pair to form,  $u_k$ is fixed through the normalization
$|u_k|^2+|v_k|^2=1$ for every $k$.
\begin{figure}[t]
\input{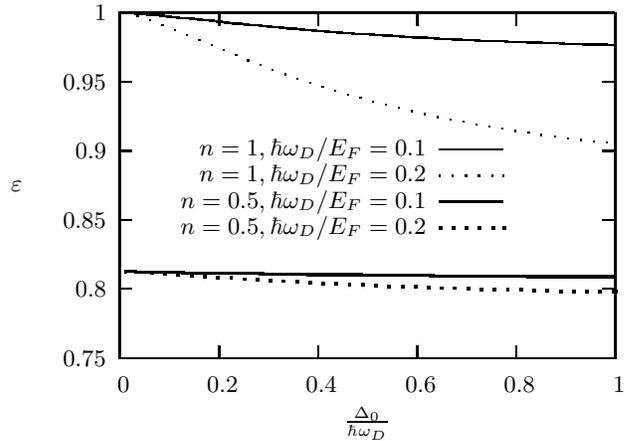}
\caption{ The entanglement in the superconducting state as a function of the energy gap $\Delta_0$,
for a few values of the electron density and the Debye frequency. The global entanglement
measure decreases for  a nonzero gap, indicating that the normal state at $\Delta_0=0$ with an
optimal double occupancy  maximizes the entanglement.  }
 \end{figure}
The Cooper pair probability is chosen as\cite{Schrieffer}
\begin{equation}
|v_k|^2={1\over 2} (1- {E_k-E_F\over \sqrt{ (E_k-E_F)^2+\Delta_k^2}})
\end{equation}
An attractive interaction in a narrow energy range near the Fermi energy $E_F$,
would imply that the gap is nonzero only in that energy range,
$\Delta_k\approx \Delta_0, |E_k-E_F|\approx \hbar\omega_{D}$,  where, $\omega_D$
is the Debye frequency or a suitable energy scale in general superconductors. 


The signle-site RDM is different from the displayed one in Eq.4, since the number of particles
is not a conserved in this state, and is given by
\begin{equation}
\rho=\left(
\begin{array}{cccc}
1-n+d  & \zeta&0&0\\
\zeta & d&0&0\\
0 & 0&{n\over 2}-d&0\\
0 &0& 0&{n\over 2}-d
\end{array}\right).
\end{equation}
In the above the off-diagonal matrix element is related to the probability of pairing
on a given site, 
and can be written in terms of the Cooper pair amplitudes as, 
$\zeta\equiv \langle c_{i\downarrow}c_{i\uparrow}\rangle={1\over N}\sum u_k v_{-k}.$ It is straightforward to
calculate, we have
\begin{equation}
\zeta={3n \Delta_0\over 4 E_F} Sinh^{-1}{\hbar \omega_D\over \Delta_0}.
\end{equation}
The double occupancy in this state is given by
$d={n^2\over 4}+\zeta^2$, which is larger than the optimal metallic value due to the pairing
tendency as we argued earlier.
Now, it is easy to calculate he global entanglement measure in the state, which is shown in
Fig.2
as a function of the superconducting order parameter
for a few values of $\hbar\omega_D/E_F$ and the electron density $n$.
It can
be seen from the plot, that for nonzero $\zeta$ (which means the double occupancy is
more than the optimal metallic value) the entanglement decreases, though only slightly
for the half-filled case, about 15 per cent for smaller electron filling. 

\begin{figure}[t]
\input{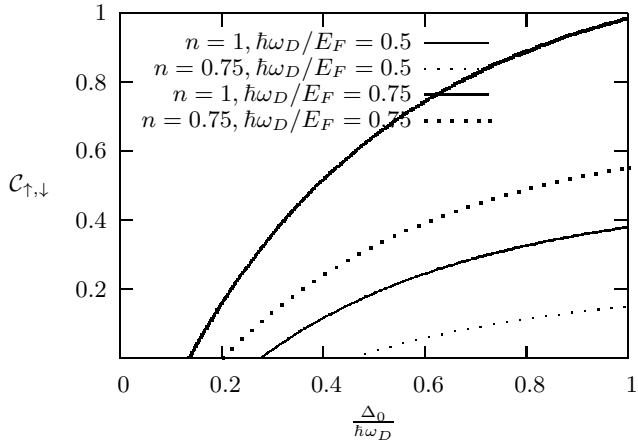}
\caption{ The on-site concurrence between up and down spins in a superconductor for different
electron densities and Debye frequencies.}
\end{figure}
The promotion of double occupancy by explicit introduction of off-diagonal
correlations can introduce an entanglement structure of different variety,  viz.
a single-site concurrence, similar to the two-site concurrence in spin-only states
\cite{Wootters}. 
Two-site concurrence in spin systems is a measure of the entanglement between two
sites in a many-spin state, viz. the bipartite entanglement between two spatial
parts labeled by the two sites. Here, in the superconducting state, the one-site
RDM can be viewed as a bipartite density matrix of two distinct spin parts. Let us
rewrite the basis states of the RDM as
$|0>=|0>_\uparrow|0>_\downarrow, |\uparrow\downarrow>=|1>_\uparrow|1>_\downarrow,
|\uparrow>=|1>_\uparrow|0>_\downarrow,|\downarrow>=|0>_\uparrow|1>_\downarrow$.
 Now the basis states have the bipartite direct-product structure of up-spin states and down-spin
states. Naturally, one can ask how much entanglement is there between the up spin
and the down spin (which are used to label the two distinct parts of the bipartite
Hilbert space). Using the Wootters' formula for he concurrence of the four-dimensional
RDM, we have the single-site concurrence in the superconducting state as
\begin{equation}
 {\cal C}_{\uparrow,\downarrow}=2 {\rm Max} (\zeta-|{n\over 2}-d|,0).
\end{equation}
 Thus, there is possibility
of a nonzero on-site concurrence, depending on the pair order parameter, or the
energy gap of the superconducting state. For example, we have  ${\cal C}_{\uparrow,\downarrow}\ne0$,
 if $\zeta(1+\zeta)\ge {n\over2}(1-{n\over2})$. We have plotted the single-site concurrence calculated
above in Fig.3 as a function of $\Delta_0/\hbar \omega_D$, for a few representative values
of the fermi energy and the electron filling.

In conclusion, we have examined the average impurity of the single-site reduced density matrix
as a global entanglement measure in many-electron systems. The maximum global entanglement occurs for the half filled uncorrelated metallic state, which optimizes densities of up and down electrons,
the holes and the doubly-occupied sites, $n_\uparrow=n_\downarrow=n_{hole}=d=1/4$.  On-site density-density repulsive interactions, or strong spin correlations, tend to inhibit
double occupancy, thus  decrease the entanglement,  as illustrated by Gutzwiller projection states.
Superconductor states promote double occupancy, over and above the optimal value of the
metallic states, but still the entanglement decreases.  We have shown that there is possibility of a nonzero on-site concurrence in the superconducting states depending on the energy gap.

  \end{document}